\documentclass[cameraready]{Interspeech}

\title{Interpretable Audio Editing Evaluation via Chain-of-Thought Difference-Commonality Reasoning with Multimodal LLMs}

\author[affiliation={1}, orcid=0009-0001-2407-0789]{Yuhang}{Jia}
\author[affiliation={1}, orcid=0009-0007-2676-0576]{Xu}{Zhang}
\author[affiliation={1}, orcid=0009-0002-6976-268X]{Yang}{Chen}
\author[affiliation={1}, orcid=0009-0003-8057-4644]{Hui}{Wang}
\author[affiliation={1}, orcid=0009-0002-3332-6675]{Enzhi}{Wang}
\author[affiliation={1,2}, orcid=0009-0000-2748-3020, correspondingauthor]{Yong}{Qin}

\address{
    $^1$ College of Computer Science, Nankai University, Tianjin, China \\
    $^2$ Academy for Advanced Interdisciplinary Studies, Nankai University, Tianjin, China
}

\email{2120240729@mail.nankai.edu.cn, qinyong@nankai.edu.cn}


\usepackage{comment}
\usepackage{multirow}

\begin{document}

\maketitle

\begin{abstract}
Automatic mean opinion score (MOS) prediction serves as a principled alternative to both subjective listening tests and objective metrics, providing scalable and consistent audio evaluation. Inspired by the LLM-as-Judge paradigm, recent multimodal large language models offer strong perceptual modeling and reasoning capabilities, enabling audio quality assessment. In this work, we address the challenging problem of audio editing evaluation and propose the first natural language-based automated evaluation framework built upon Qwen2-Audio. Two caption-based fine-tuning tasks are introduced to enhance multi-audio understanding, together with a designed Chain-of-Thought prompting strategy to encourage structured, step-by-step reasoning. Experiments show that our framework produces interpretable and logically consistent text-based evaluations, aligning closely with human judgments while outperforming existing baselines. Code and tools will be released.
\end{abstract}

\section{Introduction}
\label{sec:intro}
Human evaluation of generative models is often costly, time-consuming, and difficult to reproduce. As a scalable and reproducible alternative, automatic MOS prediction not only reduces reliance on human testing but also provides a more human-aligned perspective than traditional objective metrics. Due to these advantages, automatic MOS prediction has been widely adopted across diverse generative tasks, including image generation \cite{fekete2023vienna, jin2024paintings, jin2024apddv2}, speech generation, conversion, and enhancement \cite{cooper2022generalization, wang2023ramp, wang2023intermediate, wang2024uncertainty, tang2024singmos, mittag2021nisqa, chen2022impairment}, as well as general audio generation (e.g., sound, vocal, and music) \cite{liu2025musiceval, wang2025audioeval, yao2025songeval, jia2025towards}. Existing automated evaluation systems have demonstrated strong correlation with human subjective assessments, making them reliable tools for evaluation, model selection, and reinforcement learning.

With the rapid advancement of speech language models (SLMs) and multimodal large language models (MLLMs)\cite{chu2023qwen, chu2024qwen2, xu2025qwen2, wang2411freeze, ding2025kimi, lu2025desta2, zeng2024glm}, their enhanced perceptual and reasoning abilities have broadened the scope of automatic quality assessment. Recent studies increasingly leverage these models for audio quality evaluation, moving beyond traditional scalar or low-dimensional quality scores to more complex and holistic dimensions such as aesthetics, creativity, fairness, and comparative judgment \cite{tjandra2025meta, yao2025songeval, wang2025tta, jia2025towards}. Meanwhile, the evaluation paradigm is shifting from 1–5 MOS ratings to fine-grained, text-based assessments that deliver detailed descriptions, in-depth analyses, and even suggestions \cite{you2024depicting, chen2025audio, wang2025speechllm}. Such comprehensive and interpretable evaluation not only deepens our understanding of generative models’ characteristics and limitations but also offers valuable guidance for their targeted optimization.

In this paper, we focus on \textbf{automatic audio editing}, a task that, unlike conventional text-to-audio generation, requires models to modify existing audio according to user-provided editing instructions or target audio descriptions. Typical editing operations include addition, deletion, replacement, inpainting, or super-resolution of audio events \cite{wang2023audit, paissan2023audio, xu2024prompt, manor2024zero, han2023instructme, jia2025audioeditor,  liang2024wavcraft}. Evaluating the quality of edited audio is particularly challenging, as it requires joint perception of paired audio samples along with the associated textual instructions and descriptions. Furthermore, due to the lack of a supervised audio editing benchmark, conventional TTA metrics are difficult to apply for measuring editing quality, as they rely on ground-truth references that are not available in this setting. Consequently, a comprehensive editing evaluation must consider at least three aspects: (1) the overall quality of the edited output relative to the original audio, (2) the extent to which the output satisfies the intended editing instructions, and (3) the preservation of unedited regions and the overall characteristics of the original audio \cite{jia2025towards, jia2025audioeditor}.

\begin{figure*}[t]
\centerline{\includegraphics[width=0.95\textwidth]{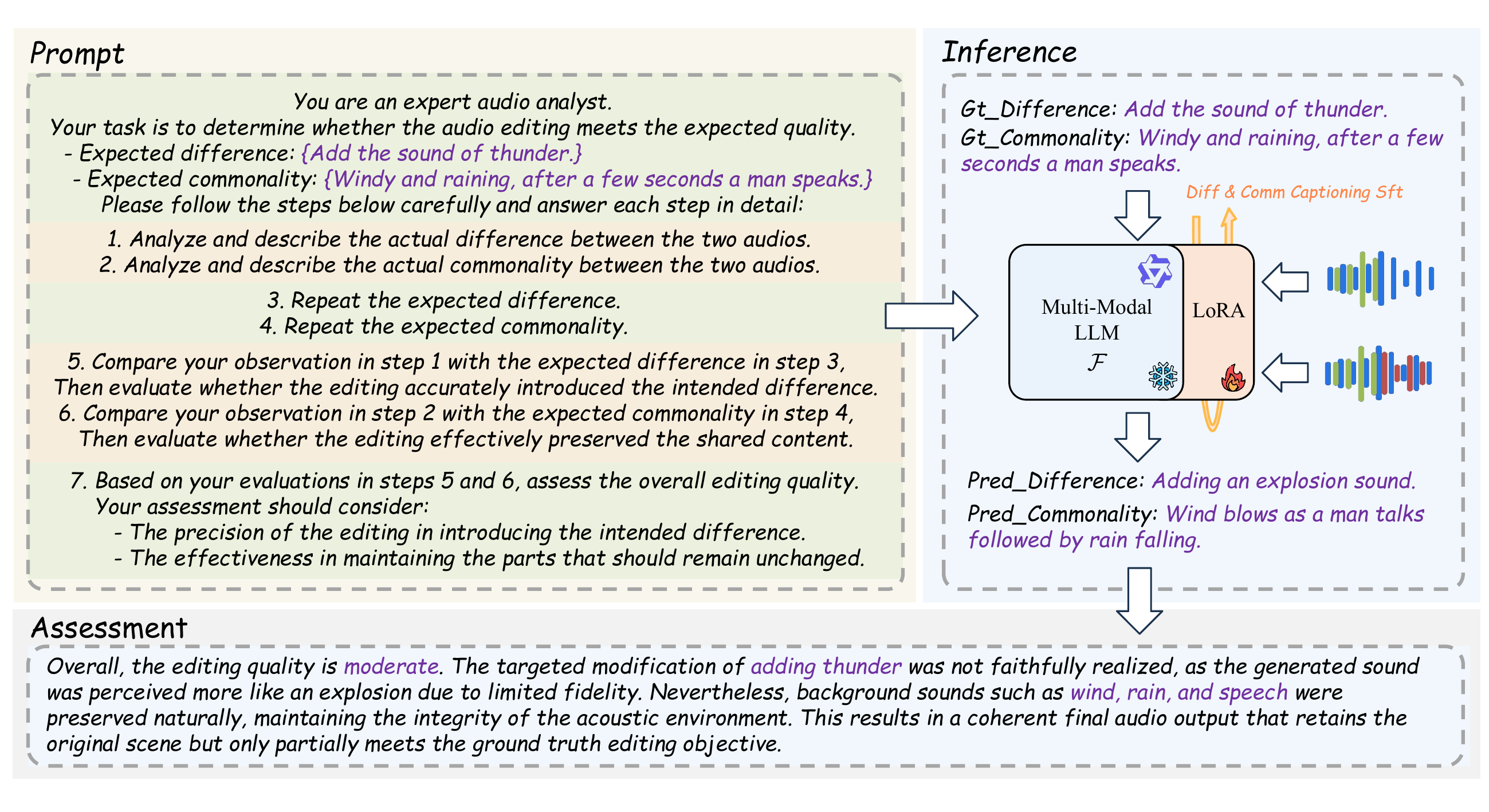}}
\vspace{-5mm}
\caption{Overview of our audio editing evaluation framework. Equipped with a 7-step CoT prompting strategy and fine-tuned MLLMs, the framework generates accurate and interpretable text-based evaluation results from complementary perspectives. \vspace{-5mm}} 
\label{fig1}
\end{figure*}

To address these challenges, we build upon Qwen2-Audio \cite{chu2024qwen2}, leveraging its strong audio understanding capability. We further improve its cross-audio reasoning through specialized caption-based fine-tuning tasks, together with carefully designed Chain-of-Thought (CoT) instruction tuning. Based on these enhancements, we develop a natural language–based framework for automated audio editing evaluation, with the following key contributions:

\textbf{1) Boosting multi-audio joint understanding.} We enhance Qwen2-Audio via two dedicated caption-based fine-tuning tasks, including Audio Difference Captioning and Audio Commonality Captioning, which significantly improve its cross-audio understanding capability. These tasks enable the model to simultaneously perceive detailed differences and shared characteristics between paired audio samples, forming the foundation for reliable editing evaluation.

\textbf{2) Correlation Validation with Human Perception.} We systematically evaluate the accuracy of the generated difference and commonality analyses by examining their correlations with conventional objective metrics and human ratings. The results demonstrate strong logical consistency and alignment with human perception. Based on this validation, we propose two novel caption-based scoring metrics, Edit\_score and Faith\_score, which achieve competitive correlations compared to traditional 1–5 scale MOS prediction models specifically adapted for audio editing.

\textbf{3) Interpretable Audio Editing Evaluation.} We introduce a 7-step Chain-of-Thought (CoT) prompting strategy combined with lightweight instruction tuning. The framework incorporates evaluation-oriented mechanisms, including attention-leakage minimization and reference repeat, to ensure structured reasoning and stable cross-audio comparison. These designs enable interpretable, text-based automated audio editing evaluation and lead to substantial improvements in logical consistency and assessment reliability.

\vspace{-1mm} 
\section{Method}
\label{sec:format}

\begin{figure*}[t]
\centerline{\includegraphics[width=1.0\textwidth]{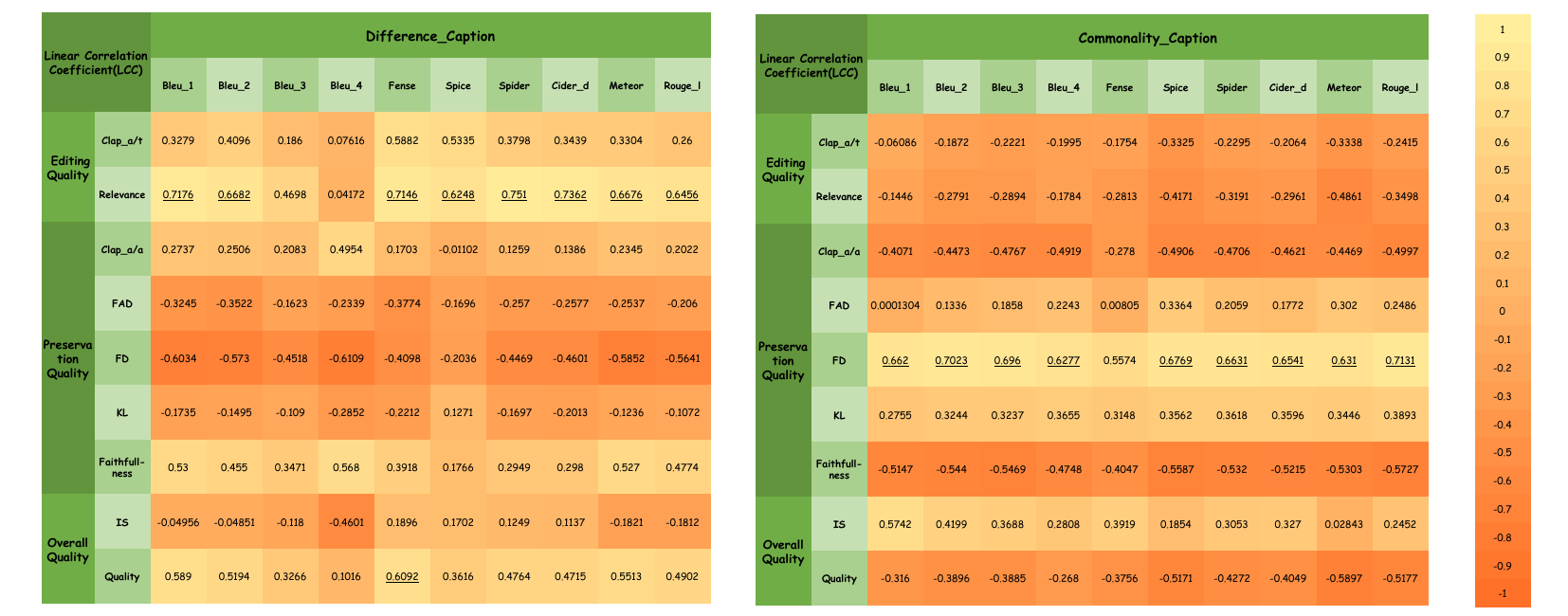}}
\vspace{-2mm}
\caption{Linear correlation coefficient (LCC) analysis between the model’s predicted Difference/Commonality accuracy and subjective/objective editing metrics across 23 audio editing systems, exhibit fully complementary correlation trends. \vspace{-5mm}} 
\label{fig2}
\end{figure*}

\subsection{Problem Formulation}
We formulate the audio editing evaluation task as follows. Given a pre-edit audio segment $A_{\text{orig}}$, its original description $T_{\text{orig}}$, a target description or editing instruction $T_{\text{target}}$, and the post-edit audio $\hat{A}_{\text{edit}}$ generated by an editing model, the goal is to provide a comprehensive text-based evaluation $E$. Specifically, $E$ comprises at least three aspects: \textbf{1) $E_o$: Overall} quality of the edited audio, \textbf{2) $E_e$: Editing} effectiveness, i.e., how accurately the target modifications were applied, and \textbf{3) $E_p$: Preservation} of unedited content and characteristics from the original audio.
Formally, the evaluation framework can be expressed as a function $\mathcal{F}$ that maps the inputs to the evaluation results:
\vspace{-2mm}
\begin{multline}
E = \mathcal{F}(A_{\text{orig}}, T_{\text{orig}}, T_{\text{target}}, \hat{A}_{\text{edit}}), \quad
E = \{E_o, E_e, E_p\}.
\end{multline}

\subsection{Fine-Tuning for Multi-Audio Joint Understanding}
\label{sec:ft}
\vspace{-1mm}
While MLLMs such as Qwen2-Audio are pretrained on audio captioning tasks and thus possess strong single-audio semantic understanding, their ability for multi-audio reasoning remains limited. To address this, we construct a paired dataset of 30,000 edited audio samples following \cite{wang2023audit, jia2025towards}. Each pair consists of an original audio clip and its edited counterpart, with the corresponding editing instruction serving as the \textit{target difference} and the overlapping content serving as the \textit{target commonality}. For the \textit{target commonality}, we handle three types of editing operations: for \textbf{addition}, the commonality caption is the original caption; for \textbf{deletion}, it is the edited caption; and for \textbf{replacement}, it is the intersection of the two captions.

We then jointly fine-tune the model to predict both the differences (difference captioning) and the preserved parts (commonality captioning) of paired edited audios. This approach requires only a modest dataset size while substantially enhancing Qwen2-Audio’s multi-audio reasoning capability, building upon its existing perceptual understanding.
\vspace{-2mm}
\subsection{Correlation Validation with Human Perception}
\vspace{-1mm}
While the model is capable of generating Difference and Commonality descriptions, it is essential to verify whether these predicted texts faithfully reflect actual editing and preservation performance. To this end, we conduct a correlation study using the AuditScore \cite{jia2025towards} audio editing evaluation dataset, which covers 23 audio editing systems and contains 6,300 annotated instances. Each instance consists of paired audio samples with textual descriptions and editing instructions, and is rated on a 1–5 scale by expert annotators from three subjective perspectives: overall quality (\textbf{Quality}), editing effectiveness (\textbf{Relevance}), and preservation of unedited content (\textbf{Faithfulness}). In addition, we compute a set of widely used objective audio editing metrics for each instance. These subjective and objective indicators are further grouped into three categories: editing ability, preservation ability, and overall quality. Finally, we calculate the linear correlation coefficient between the predicted Difference and Commonality accuracy (measured by captioning metrics) and the corresponding subjective and objective evaluation metrics, as illustrated in Figure~\ref{fig2}.

The results in Figure~\ref{fig2} reveal complementary correlation patterns between Difference-captioning and Commonality-captioning with respect to editing effectiveness and preservation. Specifically, Difference-captioning is more closely associated with editing performance, whereas Commonality-captioning is more related to preservation. This observation is consistent with the intuitive roles of differences and commonalities in audio editing. Based on this finding, we select the captioning metrics that exhibit the strongest correlations with editing effectiveness and preservation, respectively, and combine them to construct two composite metrics: \textit{Edit\_score} and \textit{Faith\_score}. These metrics leverage the complementary information from Difference-captioning and Commonality-captioning to provide quantitative evaluation of editing effectiveness and preservation: 
{
\begin{multline}
\text{Edit\_score} = \text{sigmoid} (
\log ( w_f \, D_{\text{Fense}} + w_s \, D_{\text{Spider}} + \\
\varepsilon \cdot \exp(-\lambda_{\text{edit}} \cdot (v_{sp} \, C_{\text{Spice}} + v_{me} \, C_{\text{Meteor}})) )
),
\end{multline}
}
\vspace{-6mm}
{
\begin{multline}
\text{Faith\_score} = -\text{sigmoid} (
\log ( u_{sp} \, C_{\text{Spice}} + u_{rl} \, C_{\text{RougeL}} + \\
\varepsilon \cdot \exp(-\lambda_{\text{faith}} \cdot (z_{sp} \, D_{\text{Meteor}} + z_{me} \, D_{\text{RougeL}})) )
),
\end{multline}
}
where $D\_$ and $C\_$ denote the Difference and Commonality caption metrics, respectively, and the weights are determined in proportion to the metric’s LCC values : $w_f=0.48$, $w_s=0.52$, $v_{sp}=0.46$, $v_{me}=0.54$, $u_{sp}=0.48$, $u_{rl}=0.52$, $z_{sp}=0.53$, $z_{me}=0.47$, $\varepsilon=1e^{-6}$, $\lambda_{\text{edit}}=0.5$, and $\lambda_{\text{faith}}=0.5$.

\begin{table*}[h!]
\centering
\caption{Performance comparison of the base MLLM before and after SFT on the Difference/Commonality captioning tasks. \vspace{-1mm}}
\renewcommand{\arraystretch}{0.9}
\label{tab1}
\begin{tabular}{cccccccccccc}
\toprule
\multirow{3}{*}{Model} & \multicolumn{10}{c}{Difference\_Caption↑} \\
\cmidrule(lr){2-11}
& Bleu\_1 & Bleu\_2 & Bleu\_3 & Bleu\_4 & Fense & Spice & Spider & Cider\_d & Meteor & Rouge\_l \\
\midrule
Qwen2-Audio & 0.0392 & 0.0116 & 0.0020 & 0.0000 & 0.2633 & 0.0581 & 0.0302 & 0.0024 & 0.0397 & 0.0435 \\
Qwen2-Audio+sft & \textbf{0.7542} & \textbf{0.6226} & \textbf{0.5613} & \textbf{0.5219} & \textbf{0.8370} & \textbf{0.5701} & \textbf{2.8121} & \textbf{5.0542} & \textbf{0.4023} & \textbf{0.7113} \\
\midrule
\multirow{3}{*}{Model} & \multicolumn{10}{c}{Commonality\_Caption↑} \\
\cmidrule(lr){2-11}
& Bleu\_1 & Bleu\_2 & Bleu\_3 & Bleu\_4 & Fense & Spice & Spider & Cider\_d & Meteor & Rouge\_l \\
\midrule
Qwen2-Audio & 0.0711 & 0.0242 & 0.0089 & 0.0032 & 0.2664 & 0.0267 & 0.0250 & 0.0232 & 0.0459 & 0.0615\\
Qwen2-Audio+sft & \textbf{0.3264} & \textbf{0.2095} & \textbf{0.1323} & \textbf{0.0831} & \textbf{0.6929} & \textbf{0.2273} & \textbf{0.5638} & \textbf{0.9003} & \textbf{0.1766} & \textbf{0.3216} \\
\bottomrule
\vspace{-6mm}
\end{tabular}
\end{table*}

\vspace{-1mm}
\subsection{Chain-of-Thought Prompt and Instruction Tuning}
\vspace{-1mm}
To enable natural language–based audio editing evaluation, we design a 7-step CoT prompting strategy, as illustrated in Figure~\ref{fig1}. First, the model is prompted to generate the \textit{Difference} and \textit{Commonality} descriptions for the input audio pair (pre- and post-edit), leveraging the fine-tuned multi-audio reasoning capability introduced in Section~\ref{sec:ft}. Second, it is prompted to restate the expected \textit{Difference} and \textit{Commonality} captions derived from the editing instructions. Next, the model compares the generated and expected Difference/Commonality descriptions to produce separate evaluations of editing effectiveness and preservation ability. Finally, it synthesizes these aspect-level evaluations into a comprehensive assessment, providing detailed analysis and suggestions for improvement.

During instruction fine-tuning, directly providing the expected Difference and Commonality captions while requiring the model to predict them may introduce attention leakage, causing the model to trivially copy the ground-truth captions instead of performing genuine prediction, which can lead to catastrophic forgetting. To mitigate this issue, we randomly shuffle the ground-truth captions within each batch during fine-tuning to intentionally decouple them from the corresponding inputs, thereby \textbf{reducing attention leakage} between inputs and outputs. Furthermore, we observe that the model may forget or override the expected captions after generating its own predictions during inference. To address this, we introduce two intermediate steps before the comparison stage, prompting the model to explicitly restate the standard captions in order to \textbf{reinforce reference information}. Finally, we construct a small yet high-quality training set of 40 samples through LLM-assisted generation followed by manual refinement. This set is rewritten by Qwen2-Audio and used for lightweight instruction tuning to \textbf{enhance task-specific instruction-following capability}.

\section{Experiments and Results}
 
\begin{table}[t!]
\centering
\caption{Correlation validation results. Higher values indicate stronger alignment with human judgments. \vspace{-1mm}}
\label{tab:performance_comparison}
\renewcommand{\arraystretch}{1.0}
\setlength{\tabcolsep}{1.9pt}
\begin{tabular}{ccccccc}
\toprule
\multirow{2}{*}{Model} & \multicolumn{2}{c}{Lcc↑} & \multicolumn{2}{c}{Srcc↑} & \multicolumn{2}{c}{Katu↑} \\
\cmidrule(lr){2-3} \cmidrule(lr){4-5} \cmidrule(lr){6-7}
 & Edit. & Presv. & Edit. & Presv. & Edit. & Presv. \\
\midrule
AuditEval-ssl & 0.6196 & \textbf{0.8460} & 0.6472 & \textbf{0.8809} & 0.4783 & \textbf{0.7024} \\
Edit\_score & \underline{\textbf{0.7652}} & 0.3799 & \underline{\textbf{0.7312}} & 0.3501 & \underline{\textbf{0.8460}} & 0.3320 \\
Faith\_score & 0.4260 & \underline{0.5908*} & 0.1532 & \underline{0.4605} & 0.0988 & \underline{0.3518} \\
\bottomrule
\vspace{-8mm}
\end{tabular}
\normalsize
\end{table}

\subsection{Experiment Setup and Details}
\textbf{Multi-Audio Captioning Fine-tuning.} We adopt Qwen2-We adopt Qwen2-Audio-7B\footnote{\url{https://huggingface.co/Qwen/Qwen2-Audio-7B}}
 as the backbone MLLM, and fine-tune it on 30,000 pseudo-paired audio editing samples constructed via mixing and rearrangement, following the methodology of \cite{jia2025towards}. The dataset is split into training, validation, and test sets with an 8:1:1 ratio. For evaluation, we employ standard captioning metrics, including BLEU-1–4, FENSE, SPICE, SPIDER, CIDEr\_d, METEOR, and ROUGE-L\footnote{\url{https://github.com/Labbeti/aac-metrics}}
. All experiments are conducted on four NVIDIA GeForce RTX 4090 GPUs. We fine-tune the Qwen2-Audio-7B-Instruct model using LoRA (rank = 8, $\alpha$ = 32, dropout = 0.05), with an effective batch size of 16, a maximum input length of 2048 tokens, and a learning rate of $1 \times 10^{-4}$ until convergence.

\textbf{Correlation Validation Experiment.} We adopt the AuditScore~\cite{jia2025towards} dataset, which provides comprehensive paired audio editing results with expert 1–5 subjective ratings, supplemented by our objective measures (including Clap\_Score, FAD, FD, KL, and IS)\footnote{\url{https://github.com/haoheliu/audioldm_eval}}. For the score comparison experiment (Table~\ref{tab:performance_comparison}), we evaluate our proposed \textit{Edit\_Score} and \textit{Faith\_Score} against the baseline scoring model AuditEval-ssl, using Linear Correlation Coefficient (LCC), Spearman’s Rank Correlation Coefficient (SRCC), and Kendall’s Tau Correlation Coefficient (KTAU) as evaluation metrics.

\textbf{A/B Test Ablation and Comparative Study.} We conduct experiments on the AuditScore test set (318 samples). Qwen2.5-Omni\footnote{\url{https://huggingface.co/Qwen/Qwen2.5-Omni-7B}} is employed as the judge to cast votes based on three criteria: completeness (coverage of both editing and preservation, plus an overall assessment), accuracy (correctness of editing/preservation descriptions, identified events, and reasoning consistency), and richness (inclusion of details, logical explanations, and suggestions). We use raw Qwen2-Audio with a non-optimized prompt as a baseline, and further adopt Qwen2.5-Omni(the judger) with optimized prompt as a strong comparative baseline against our method. During the A/B tests, the order of the two options is alternated to eliminate potential position bias of the voting model.

\vspace{-1mm}
\subsection{Results and Analysis}
\vspace{-1mm}
\textbf{Multi-Audio Captioning Fine-tuning.} As shown in Table~\ref{tab1}, although Qwen2-Audio is pre-trained on more than 10k hours of general audio data, it shows almost no capability in joint multi-audio captioning due to the lack of specialized training tasks design. After fine-tuning on only about 80 hours of our curated paired data, its ability to perform Difference and Commonality captioning is significantly boosted, achieving strong performance with FENSE scores of 0.83 and 0.69, respectively.

\textbf{Correlation Validation Experiment.} As shown in Fig.~\ref{fig2}, the model-predicted accuracy of Difference Captions for pre- and post-edit audio exhibits a strong positive correlation with objective and subjective metrics of editing effectiveness, e.g., Clap\_at/FENSE = 0.58 and Relevance/FENSE = 0.71, while showing generally negative correlations with preservation-related metrics, e.g., FD/METEOR = -0.58 and FAD/FENSE = -0.37. In contrast, the accuracy of Commonality Captions demonstrates a strong positive correlation with preservation ability, e.g., FD/SPICE = 0.67 and FD/ROUGE\_L = 0.71, but a negative correlation with editing effectiveness metrics, e.g., Relevance/SPICE = -0.41 and Clap\_at/METEOR = -0.33. This fully complementary correlation pattern and its logical consistency validate the feasibility of leveraging the model’s Difference/Commonality-telling capabilities for editing evaluation, providing support for subsequent experiments.

\begin{figure}[t]
\centerline{\includegraphics[width=0.48\textwidth]{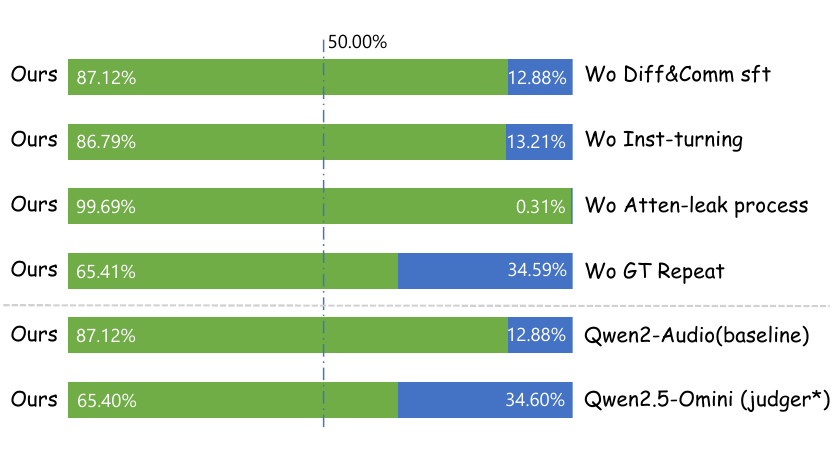}}
\vspace{-2mm}
\caption{Ablation Results (top) and Comparison with the baseline raw model and A/B test judger model (bottom).}
\vspace{-5mm}
\label{fig3}
\end{figure}

Moreover, as shown in Table~\ref{tab:performance_comparison}, our proposed combined-score metric \textit{Edit\_score}, the first objective audio editing metric computed using an MLLM, surpasses the specialized supervised 1–5 scoring model \textit{AuditEval-ssl} in predicting editing effectiveness, aligning more closely with human expert ratings and demonstrating the potential of leveraging unsupervised LLMs for complex automatic evaluation tasks. The \textit{Faith\_score} is less effective in predicting preservation ability compared to \textit{AuditEval-ssl}, as the AuditScore benchmark, in addition to considering semantic consistency, places substantial emphasis on acoustic consistency, including prosody, volume, and noise, which may pose greater challenges for MLLMs to capture. Nevertheless, by combining multiple metrics, \textit{Faith\_score} achieves a correlation of 0.59 with the human-rated faithfulness metric, outperforming any constituent metric alone.

\textbf{A/B Test Ablation and Comparative Study.} As shown in Fig.~\ref{fig2}, removing Difference/Commonality caption fine-tuning, the inclusion of 40 carefully curated assessment one-shot cases, or instruction-tuning strategies such as the attention-leakage mitigation process and ground-truth repetition, severely degrades the model’s text-based automated subjective evaluation capability. Notably, without the attention-leakage mitigation process, the model, when trained with correct answers, tends to confuse the captioning task with simple repetition, which results in catastrophic forgetting and incoherent captioning outputs. In comparative experiments against both the raw Qwen2-Audio model without CoT prompting and Qwen2.5-Omni, our framework demonstrates clear superiority, highlighting the effectiveness of the proposed automatic interpretable, natural language–based audio editing evaluation approach.

\section{Conclusion}
In this work, we present an interpretable, natural language–based framework for automated audio editing evaluation, built upon Qwen2-Audio with Difference- and Commonality-focused fine-tuning and CoT instruction tuning. By employing carefully designed strategies, our approach delivers rich, interpretable, and descriptive assessments that closely align with both human judgments and objective metrics. These results provide a foundation for reliable evaluation, informed model selection, and future reinforcement learning–based optimization of MLLM-driven audio editing systems.

\section{Acknowledgments}
Supported by the National Key R\&D Program of China (Grant No.2022ZD0116307) and NSF China
(Grant No.62271270).

\section{Generative AI Use Disclosure}
During manuscript preparation, generative AI tools were used solely to improve language clarity and readability. All content was reviewed and verified by the authors, who take full responsibility for its accuracy and originality. Notably, we also employed LLMs as judges for text-based evaluation, but no generative AI was used to produce substantive scientific content.

\bibliographystyle{IEEEtran}
\bibliography{mybib}

\end{document}